\documentclass[prc,superscriptaddress,unsortedaddress,twocolumn,showpacs,preprintnumbers,amsmath,amssymb,floatfix]{revtex4}
\usepackage[dvips]{graphicx}
\usepackage{amsmath}
\usepackage{amssymb}
\usepackage{mathrsfs}
\usepackage{multirow}
\usepackage{bm}
\usepackage{color}

\def\la{\mathrel{\mathpalette\fun <}}
\def\ga{\mathrel{\mathpalette\fun >}}
\def\fun#1#2{\lower3.6pt\vbox{\baselineskip0pt\lineskip.9pt
\ialign{$\mathsurround=0pt#1\hfil##\hfil$\crcr#2\crcr\sim\crcr}}}

\def\a{\alpha}

\newcommand{\beq}{\begin{equation}}
\newcommand{\eeq}{\end{equation}}
\newcommand{\bea}{\begin{eqnarray}}
\newcommand{\eea}{\end{eqnarray}}

\newcommand{\bfi}[1]{\mbox{\boldmath $#1$}}

\newcommand{\vs}{{\bfi s}}
\newcommand{\vt}{{\bfi t}}
\newcommand{\vT}{{\bfi T}}

\newcommand{\vrr}{{\bfi r}}
\newcommand{\vR}{{\bfi R}}

\newcommand{\vX}{{\bfi X}}

\begin{document}


\title{Mass-number and isotope dependence of \\
local microscopic optical potential for polarized proton scattering}

\author{Masakazu Toyokawa}
\email[]{toyokawa@phys.kyushu-u.ac.jp}
\affiliation{Department of Physics, Kyushu University, Fukuoka 812-8581, Japan}
,
\author{Kosho Minomo}
\email[]{minomo@phys.kyushu-u.ac.jp}
\affiliation{Department of Physics, Kyushu University, Fukuoka 812-8581, Japan}
,
\author{Masanobu Yahiro}
\email[]{yahiro@phys.kyushu-u.ac.jp}
\affiliation{Department of Physics, Kyushu University, Fukuoka 812-8581, Japan}

\date{\today}

\begin{abstract}
We derive local microscopic optical potentials $U$ systematically 
for polarized proton scattering at 65~MeV using 
the local-potential version of the Melbourne $g$-matrix folding model. 
As target nuclei, we take $^{6}$He and neutron-rich Ne isotopes 
in addition to stable nuclei of mass number $A=4$--$208$ in order to 
clarify mass-number and isotope dependence of $U$. 
The local potentials reproduce the experimental data systematically 
and have geometries similar to the phenomenological optical potentials for 
stable targets. 
The target density is broadened by the weak-binding nature and/or 
deformation of unstable nuclei. 
For the real spin-orbit part of $U$ 
the density broadening weakens the strength and enlarges the radius, 
whereas for the central part it enlarges both of the strength and 
the radius. 
The density-broadening effect is conspicuous for halo nuclei such 
as $^{6}$He and $^{31}$Ne. 
Similar discussions are made briefly for proton scattering at 200~MeV. 
We briefly investigate how the isovector 
and the non spherical components 
of $U$ affect proton scattering. 
\end{abstract}

\pacs{25.40.Cm, 24.10.Ht, 24.70.+s}

\maketitle
\section{Introduction}

Understanding of nucleon-nucleus ($NA$) and nucleus-nucleus ($AA$) 
interactions is one of the most important subjects in nuclear physics. 
The interactions (optical potentials) are necessary to describe not only 
elastic scattering but also more complicated reactions. 
Actually, the interactions are inputs 
of theoretical calculations, such as 
the distorted-wave Born approximation (DWBA) 
and the continuum discretized coupled-channels method (CDCC) 
\cite{CDCC-review1,CDCC-review2,Yahiro:2012tk}, to 
analyze inelastic scattering and transfer and breakup reactions.

In general, $NA$ elastic scattering is less absorptive and, 
hence, more sensitive to nuclear interior than $AA$ scattering. 
Furthermore, one can determine not only 
the central part but also the spin-orbit part of the $NA$ interaction, 
when the incident particle is polarized. 
Systematic measurements of polarized proton ($p$) elastic scattering 
have been made for stable target nuclei. 
As an outcome of the measurements, some 
global $NA$ optical potentials have been constructed 
phenomenologically; see for example Refs.~\cite{Perey-Perey,Dirac1,Dirac2,Koning-Delaroche}. In most of the cases, 
the potentials are assumed to be local and thereby 
quite practical in many applications such as DWBA and CDCC calculations.

Elastic scattering of a polarized proton on unstable nuclei 
are a quite good tool used to investigate not only 
nuclear structures of unstable nuclei  but also 
interactions between a proton and unstable nuclei, 
but the measurements are not easy. 
The first measurement on the vector analyzing power $A_y$
was made by Uesaka and his collaborators 
for $p$+$^{6}$He elastic scattering at 71~MeV
with the inverse kinematics~\cite{Uesaka:2010}. 
The $g$-matrix folding model yields reasonable agreement with 
the measured vector analyzing power~\cite{Uesaka:2010}. 
The nuclear-medium effect included in the model 
is thus important for the observable. 
Recently, the effects of $^{6}$He breakup on the elastic scattering  
were investigated with the eikonal approximation~\cite{Kaki:2012hr}. 
The medium effect is more significant than the breakup effect, as far as 
$A_y$ is concerned.

The $g$-matrix folding model is a powerful tool to describe 
$NA$ and $AA$ interactions~\cite{M3Y,JLM,Brieva-Rook,Satchler-1979,Satchler,CEG,
Rikus-von Geramb,Amos,CEG07,rainbow}. 
Particularly when the Melbourne $g$-matrix nucleon-nucleon ($NN$) 
interaction~\cite{Amos} 
is used, the model is successful in reproducing polarized 
proton elastic scattering from stable target nuclei systematically 
with no free parameter~\cite{Amos}. 
In the model, target-excitation and projectile-breakup effects are taken 
into account within the local-density approximation in addition to 
the medium effect. 
The microscopic optical potential derived by the model is nonlocal and 
thereby not so practical in many applications. 
It is, however, possible to localize the potential with the Brieva-Rook 
approximation \cite{Brieva-Rook}. Recently 
the validity of the approximation was shown in Ref. \cite{Minomo:2009ds}.

From a theoretical viewpoint based on the multiple scattering 
theory~\cite{Watson,KMT,Yahiro-Glauber}, 
a multiple $NN$ scattering series 
in $AA$ collisions~\cite{Yahiro-Glauber} is more complicated than that in 
$NA$ collisions~\cite{Watson,KMT}. 
In this sense, microscopic understanding of $AA$ interaction 
is relatively more difficult than that of $NA$ interaction. 
Very recently, however, for $AA$ elastic scattering 
the local microscopic optical potential 
was derived from the Melbourne $g$-matrix interaction by using 
both the folding model 
and the Brieva-Rook localization~\cite{Min11,Min12,Sum12}. 
When the local-potential version of the Melbourne $g$-matrix folding model 
was applied to $^{12}$C+$^{12}$C scattering at 6.2, 
135, and 250 MeV/nucleon and $^{20-32}$Ne+$^{12}$C scattering 
at 240 MeV/nucleon, the local potentials reproduced the measured 
elastic cross sections and reaction cross sections $\sigma_{\rm R}$ 
with no free parameter~\cite{Min11,Min12,Sum12}. 
In the calculations, densities of unstable nuclei $^{20-32}$Ne 
were evaluated by use of antisymmetrized molecular 
dynamics (AMD)~\cite{Kimura,Kimura1} 
with the Gogny-D1S interaction~\cite{GognyD1S}, 
whereas the phenomenological density~\cite{C12-density} 
was taken for $^{12}$C. 
The AMD wave functions successfully describe low-lying
spectra of Ne isotopes~\cite{Kimura}. 
The microscopic approach shows that 
$^{30-32}$Ne in the ``island of inversion" have large deformation and 
particularly $^{31}$Ne is a deformed halo nucleus 
with spin-parity $3/2^{-}$~\cite{Min11,Min12,Sum12}. 
The local-potential version of the Melbourne $g$-matrix folding model 
is thus a powerful tool to determine structures of unstable nuclei. 
The success of the local-potential version 
of the Melbourne $g$-matrix folding model for 
proton scattering from stable nuclei and 
$^{20-32}$Ne+$^{12}$C scattering implies that 
the model is reliable also for proton scattering 
from unstable nuclei such as Ne isotopes.

In this paper, we derive local microscopic optical potentials $U$ 
systematically for polarized proton scattering at 65~MeV, 
using the local-potential version of the Melbourne $g$-matrix folding model. 
This local nature of $U$ makes it easy to clarify global properties 
of $U$ for both stable and unstable targets. 
As the targets, we consider $^{6}$He and neutron-rich Ne isotopes 
in addition to stable nuclei of mass number $A=4$--$208$. 
Throughout the global analyses, we clarify $A$ and isotope dependence 
of $U$. 
Target densities are broadened much in $^{6}$He and neutron-rich Ne isotopes 
compared with stable nuclei with the same $A$ by 
the weak-binding nature and/or the nuclear deformation. 
We clarify how the density broadening affects the shape and 
the strength of $U$. 
This analysis is interesting particularly for $^{6}$He and $^{31}$Ne, 
since $^{6}$He is a typical two-neutron halo nucleus and $^{31}$Ne is 
a one-neutron halo nucleus with large deformation. 
Similar analyses are briefly made for proton scattering at 200~MeV. 
Finally, we briefly investigate how the isovector and the non-spherical 
components of $U$ affect proton scattering.

We recapitulate the local-potential version of the Melbourne 
$g$-matrix folding model 
in Sec.~\ref{Theoretical framework} and show numerical results 
in Sec.~\ref{Results}.
Section \ref{Summary} is devoted to a summary.

\section{Theoretical framework}
\label{Theoretical framework}

Proton elastic scattering can be 
described by the one-body Schr\"odinger equation,
\bea
(T_R + U -E){\Psi}^{(+)}=0 , 
\label{schrodinger}
\eea
with a proton optical potential $U$, 
where 
$E$ is an incident energy of proton and 
$T_R$ is a kinetic-energy operator concerning the relative coordinate 
$\vR$ between an incident proton and a target. 
The optical potential $U$ can be divided into 
the central (CE), the spin-orbit (LS), and the Coulomb (Coul) component 
as follows:
\bea
U=U_{\rm CE}+U_{\rm LS} {\bfi L} \cdot {\bfi \sigma} 
+ V_{\rm Coul} . 
\eea

In the $g$-matrix folding model, $U$ consists of 
the direct and exchange parts~\cite{Arellano:1995}. The exchange part 
is nonlocal, but it can be localized with 
the Brieva-Rook approximation \cite{Brieva-Rook}. 
Validity of the approximation is shown in Ref. \cite{Minomo:2009ds}. 
We take this approximation in the present paper, since the localized $U$ 
make it possible to make systematic analyses 
of  $U$ over stable and unstable targets. 
In addition to the merit, the local microscopic optical potential is quite 
useful in many applications, particularly when 
potentials between stable and unstable nuclei are needed in the applications.

The central part $U_{\rm CE}$ is then described by the sum of 
the direct component $U_{\rm CE}^{\rm DR}$ 
and the localized exchange component 
$U_{\rm CE}^{\rm EX}$~\cite{Brieva-Rook,CEG,CEG07},
\bea
U_{\rm CE} \equiv V_{\rm CE}+iW_{\rm CE}
=U_{\rm CE}^{\rm DR}+U_{\rm CE}^{\rm EX}
\eea
with 
\bea
\label{eq:UD}
U_{\rm CE}^{\rm DR}(\vR) \hspace*{-0.15cm} &=& \hspace*{-0.15cm} 
\!\! \sum_{\alpha=p,n} \! \int \!\! \rho_{\alpha}(\vrr) 
            g^{\rm DR}_{p\alpha}(s;\rho_{\alpha}) d \vrr, \\
\label{eq:UEX}
U_{\rm CE}^{\rm EX}(\vR) \hspace*{-0.15cm} &=& \hspace*{-0.15cm} - \!\! \sum_{\alpha=p,n} \!
\int \!\! \rho_{\alpha}(\vrr,\vrr-\vs)
            g^{\rm EX}_{p\alpha}(s;\rho_{\alpha}) 
            j_0(K(R) s)
            d \vrr ,~~~~
            \label{U-EX}
\eea
where $V_{\rm CE}$ ($W_{\rm CE}$) is the real (imaginary) part of $U_{\rm CE}$ 
and $\vs=\vrr - \vR $, and $\vrr$ is the coordinate of an interacting nucleon 
from the center-of-mass (c.m.) of target. 
The local momentum $\hbar K(R) \equiv \sqrt{2\mu_{\!R}^{} (E - U_{\rm
CE}-V_{\rm Coul})}$ present in 
Eq.~\eqref{U-EX} is obtained self-consistently, where $\mu_{\! R}^{}$ is 
the reduced mass of the proton+target system.

Usually the direct part $g^{\rm DR}_{p\alpha}$ and the exchange part $g^{\rm EX}_{p\alpha}$ of the $g$-matrix interaction are assumed to be a function 
of the local density $\rho_{\alpha}=\rho_{\alpha}(\vrr-\vs/2)$
at the midpoint of the interacting nucleon pair.
The direct and exchange parts are described as
\bea
&g_{pp}^{\rm DR,EX}(s;\rho_p) = \displaystyle\frac{1}{4} \left( \pm g^{01} + 3
 g^{11}\right) ,\\
 &g_{pn}^{\rm DR,EX} (s;\rho_n) = \displaystyle\frac{1}{8} \left( g^{00} \pm
 g^{01} \pm 3 g^{10} + 3 g^{11}\right) ,
\eea
in terms of the spin-isospin component $g^{ST}$ of the $g$-matrix interaction. 


Similar derivation is possible for the spin-orbit part 
$U_{\rm LS}$~\cite{CEG07}, 
\bea
U_{\rm LS} \equiv V_{\rm LS}+iW_{\rm LS}
=U_{\rm LS}^{\rm DR}+U_{\rm LS}^{\rm EX}
\eea
with 
\bea
&& U_{\rm LS}^{\rm DR} (\vR) = \! - \frac{1}{4 R^2} \!\!\! \sum_{\alpha=p,n} \! \int \!\! \vR \! \cdot
\! \vs \rho_{\alpha} (\vrr) g^{\rm DR}_{{\rm LS},p\alpha}(s;\rho_{\alpha})d\vrr ,\\
&& U_{{\rm LS}}^{\rm EX} (\vR) = -\pi \sum_{\alpha=p,n}\int ds ~s^3 \left[
\frac{2j_0(K(R)s)}{R}\rho_1^{(\alpha)}(R,s) \right. \nonumber \\
&&~~~~~~~~~~~~~~~~~~~~~~~~~~~~~\left. + \frac{j_1(K(R)s)}{2k} \delta_0^{(\alpha)}(R,s) \right] ,
\eea 
where $V_{\rm LS}$ ($W_{\rm LS}$) is the real (imaginary) part 
of $U_{\rm LS}$ [see Appendix \ref{Definitions} for the definition of 
$\rho_1^{(\alpha)}(R,s)$ and $\delta_0^{(\alpha)}(R,s)$].
As a $g$-matrix interaction, we take the Melbourne interaction 
\cite{Amos} constructed from the Bonn-B nucleon-nucleon potential~\cite{BonnB}. 

As target densities $\rho$, we consider three kinds of matter densities. 
For all the targets analyzed here, the matter densities are evaluated 
by spherical Hartree-Fock (HF) calculations 
with the Gogny-D1S interaction~\cite{GognyD1S} in which 
the spurious center-of-mass motions are removed 
in the standard manner~\cite{Sum12}. 
For lighter stable nuclei of $A \le 40$, 
we also use the phenomenological
proton-density~\cite{C12-density} determined from electron scattering 
in which the finite-size effect of proton charge 
is unfolded in the usual way~\cite{Singhal}. 
For each nucleus with $A \le 40$, 
the neutron density is assumed to have the same geometry as the 
proton one, since the deviation of 
the neutron root-mean-square (rms) radius from the proton one 
is about 1\% in the spherical HF calculation.

For $^{20-32}$Ne, the matter densities are evaluated also by 
AMD~\cite{Kimura,Kimura1} 
with the Gogny-D1S interaction~\cite{GognyD1S}. 
This provides deformed matter densities. 
The AMD wave functions successfully describe low-lying
spectra of Ne isotopes~\cite{Kimura} and $\sigma_{\rm R}$ of 
$^{20-32}$Ne+$^{12}$C scattering at 240~MeV/nucleon~\cite{Min11,Min12,Sum12}, 
and it is free from the spurious center-of-mass motion; 
see Ref.~\cite{Kimura,Sum12} for the details of AMD calculations.

Particularly for $^{31}$Ne with a quite small one-nucleon separation energy, 
we have to do a sophisticated calculation to make a tail correction 
to the AMD density, since the AMD density is inaccurate in its tail region. 
In principle, the ground state of $^{31}$Ne can be described by
the $^{30}$Ne+$n$ cluster model with core ($^{30}$Ne) excitations, and 
one can solve coupled equations for the model with the resonating group 
method (RGM) in which the ground and excited states of $^{30}$Ne 
are constructed by AMD. This was done in Refs.~\cite{Min12,Sum12}. 
This method is referred to as AMD-RGM in this paper.

The matter density $\rho_{ImIm'}(\bm r)$ is obtainable from
the ground state wave function $\Phi^{Im\pi}_{\rm g.s.}$ with spin $I$, 
its $z$-component $m$ and parity $\pi$ as 
\begin{align}
 \rho_{ImIm'}(\bm r)&=\langle \Phi^{Im\pi}_{\rm g.s.}|\sum_{i}
 \delta(\bm r_i - \bm X - \bm r)|\Phi^{Im'\pi}_{\rm g.s.}\rangle,
 \label{eq:amddens1}\\
 &=\sum_{\lambda=0}^{2I} \rho_{II}^{(\lambda)}(r)
 (Im' \lambda \mu |Im)
 Y^*_{\lambda \mu}(\hat r),
 \label{eq:amddens}
\end{align}
where $\vX$ is the center-of-mass coordinate and $\vrr_i$ is the coordinate 
of the $i$-th nucleon, and the summation over $\lambda$
in Eq. (\ref{eq:amddens}) runs for even numbers. 
As the standard manner in the folding model, 
only the spherical ($\lambda=0$) component of $\rho_{II}^{(\lambda)}$ 
is taken. 
This approximation makes the folding potential $U$ spherical. 
This procedure is exact for even nuclei with $I=0$, but approximate 
for odd nuclei with $I \ne 0$.
The procedure is a good approximation for 
reaction cross sections of $AA$ scattering
at high incident energies~\cite{Sum12}. 
We will examine in Sec. \ref{Deformed} 
how accurate this procedure is for $NA$ scattering. 

\section{Results}
\label{Results}

In this section we analyze polarized proton scattering at $E=65$ and 200~MeV, 
using the local-potential version of the Melbourne $g$-matrix folding model, 
although the analysis is mainly focused on the case of $E=65$~MeV.  
In the folding model, we also consider 
the Franey-Love $t$-matrix $NN$ interaction~\cite{Love-Franey}. 
As target nuclei of proton scattering, we consider 
stable nuclei of $^{4}$He, $^{12}$C, $^{16}$O, $^{20}$Ne, $^{24}$Mg, $^{40}$Ca, $^{90}$Zr, $^{208}$Pb, and unstable nuclei of 
$^{6}$He and neutron-rich Ne isotopes. 

\subsection{Proton scattering from stable nuclei at 65 and 200~MeV}

In this subsection, we mainly consider proton scattering at 65~MeV and 
briefly discuss the case of 200~MeV at the end of this subsection.

\begin{figure}[htbp]
\begin{center}
 \includegraphics[width=0.4\textwidth,clip]{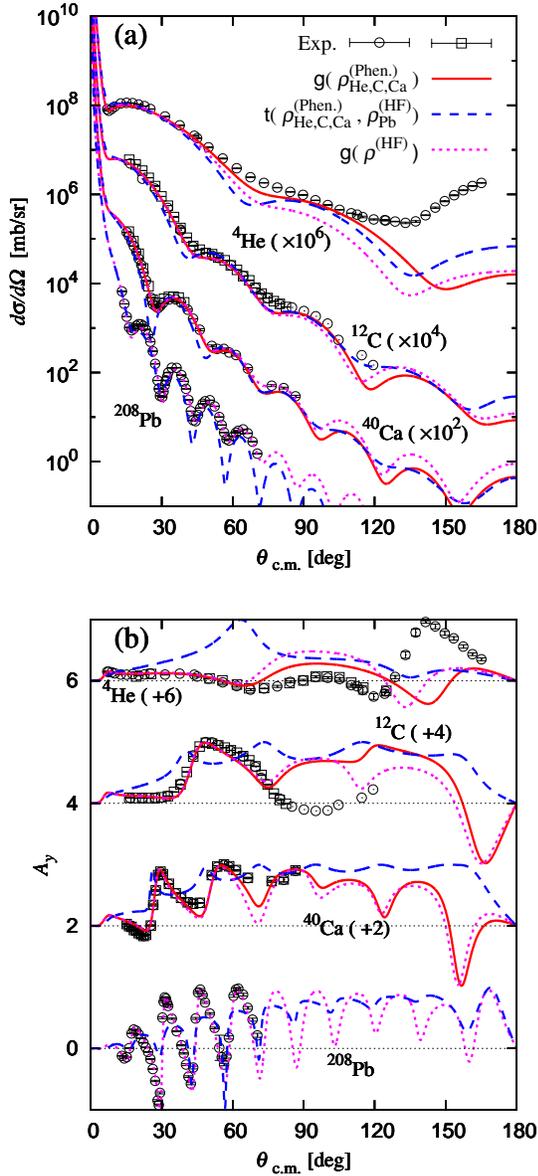}
 \caption{(Color online) Angular distributions of  
 (a) the elastic cross sections $d\sigma/d\Omega$
 and (b) the vector analyzing powers $A_y$ 
 for proton scattering at $E=65$~MeV from 
 $^{4}$He, $^{12}$C, $^{40}$Ca, and $^{208}$Pb targets. 
 The solid (dotted) lines represent the results 
 of the $g$-matrix folding model with 
 the phenomenological (HF) density, whereas 
 the dashed lines correspond to the results of 
 the $t$-matrix folding model with  the phenomenological density. 
 Experimental data are taken from Refs.~\cite{pHe4-exp1,pHe4-exp2,pC12-exp1,pC12-exp2,pCa40-exp1,pCa40-exp2,pCa40-exp3,pPb208-exp1}.
   }
 \label{Fig-Xsec-stable}
\end{center}
\end{figure}

Figure~\ref{Fig-Xsec-stable} shows elastic cross sections 
$d\sigma/d\Omega$ and vector analyzing powers $A_y$ 
as a function of the scattering angle $\theta_{\rm c.m.}$ 
for proton scattering at $E=65$~MeV from 
$^{4}$He, $^{12}$C, $^{40}$Ca, and $^{208}$Pb targets. 
The solid lines denote the results of the local microscopic potentials 
derived from the Melbourne $g$-matrix interaction. 
The results are consistent with those~\cite{Amos} 
of the nonlocal microscopic potentials 
derived from the Melbourne $g$-matrix interaction, 
although target wave functions used in two calculations differ. 
The $g$-matrix folding model shown by the solid lines 
yields better agreement with the experimental data~\cite{pHe4-exp1,pHe4-exp2,pC12-exp1,pC12-exp2,pCa40-exp1,pCa40-exp2,pCa40-exp3,pPb208-exp1} 
than the $t$-matrix folding model denoted by the dashed lines. 
The medium effect shown by the difference between the solid and dashed 
lines are more significant for $A_y$. For lighter targets 
of $^{4}$He, $^{12}$C and $^{40}$Ca, 
the agreement of the $g$-matrix folding model with the data is better 
for the phenomenological densities denoted by the solid lines than 
for the HF densities shown by the dotted lines. However, 
the $g$-matrix folding model with the HF density 
still keeps reasonable agreement with the data. 
The HF density based on the Gogny-D1S interaction~\cite{GognyD1S}
is thus quite useful to derive 
the microscopic optical potentials $U$ systematically.

\begin{figure}[htbp]
\begin{center}
 \includegraphics[width=0.45\textwidth,clip]{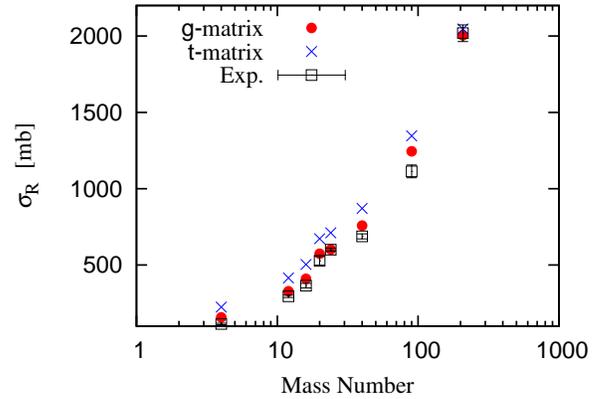}
 \caption{(Color online) $A$ dependence of the reaction cross sections 
$\sigma_{\rm R}$ for proton scattering at $E \approx 65$~MeV. 
 The closed circles (crosses) stand for the results of 
 the $g$-matrix ($t$-matrix) folding model with 
 the phenomenological densities for $A \le 40$ and the HF ones for $A > 40$. 
 Experimental data are taken 
 from Refs.~\cite{rcs-exp1,rcs-exp2,rcs-exp3,rcs-exp4,rcs-exp5}, where 
 $E=47.9$~MeV for $^{4}$He,  
 65.5~MeV for $^{12}$C,  $^{16}$O, $^{40}$Ca, and $^{208}$Pb; 
 47.0~MeV for $^{20}$Ne; 
 48.0~MeV for $^{24}$Mg; and 60.8~MeV for $^{90}$Zr. 
  }
 \label{Fig-reaction-cross-section}
\end{center}
\end{figure}

Figure~\ref{Fig-reaction-cross-section} shows $A$ dependence of 
the reaction cross sections $\sigma_{\rm R}$ for proton scattering. 
The closed circles (crosses) denote the results of 
the $g$-matrix ($t$-matrix) folding model 
with the phenomenological target densities  
for $A \le 40$ and the HF target densities  for $A > 40$. 
The $g$-matrix folding model yields better agreement with the data than 
the $t$-matrix folding model. 
The results of the $g$-matrix folding model agree with 
the data~\cite{rcs-exp1,rcs-exp2,rcs-exp3,rcs-exp4,rcs-exp5} 
within 10\% error in a wide range of $4 \le A \le 208$. 
Since $\sigma_{\rm R}$ is sensitive to $W_{\rm CE}$, 
the part is thus derived with at least 90\% accuracy.

\begin{figure}[htbp]
\begin{center}
 \includegraphics[width=0.5\textwidth,clip]{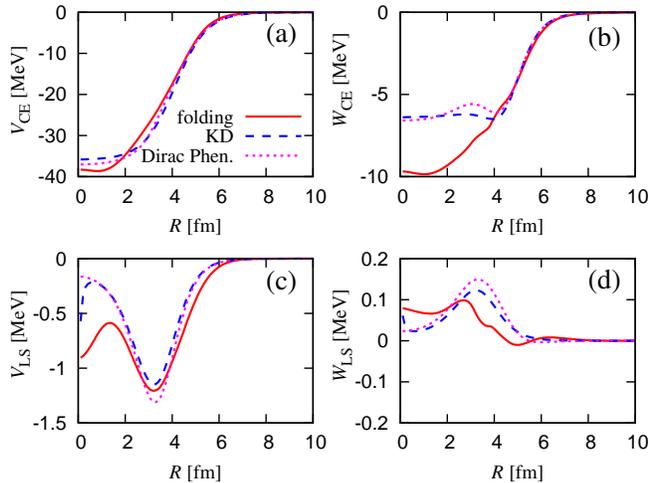}
 \caption{(Color online) 
 $R$ dependence of the microscopic optical potential, 
 the Koning-Delaroche (KD) optical potential and  the Dirac phenomenology optical potential for $p$+$^{40}$Ca scattering at $E=65$~MeV. 
 The microscopic optical potential is denoted by the solid lines, 
the Koning-Delaroche (KD) optical potential \cite{Koning-Delaroche} 
by the dashed lines and 
the Dirac phenomenology optical potential by the dotted lines. 
Panels (a), (b), (c), and (d) correspond to $V_{\rm CE}$,
 $W_{\rm CE}$, $V_{\rm LS}$, and $W_{\rm LS}$, respectively.
 }
 \label{Fig-Pot-40Ca}
\end{center}
\end{figure}

As a merit of deriving the local microscopic optical potential $U$,
we can directly compare $U$ with phenomenological optical potentials. 
Here we take two types of phenomenological optical potentials. 
One is the Koning-Delaroche (KD) optical potential \cite{Koning-Delaroche} 
based on the standard potential search 
in the framework of the Schr\"odinger equation, and another is 
the Dirac phenomenology optical potential \cite{Dirac1,Dirac2} 
based on the standard potential search in the framework of the Dirac equation. 
As an example, we compare the three optical potentials 
for $p$+$^{40}$Ca scattering at $E=65$~MeV in Fig. \ref{Fig-Pot-40Ca}. 
The microscopic optical potential is calculated from 
the phenomenological target density. 
The imaginary spin-orbit part $W_{\rm LS}$ is about 10 times as small as 
$V_{\rm LS}$, 
so it hardly affects the present elastic scattering. 
We then compare the three optical potentials for $V_{\rm CE}$, $W_{\rm CE}$ and $V_{\rm LS}$. The KD optical potential denoted by the dashed lines  
has $R$ dependence similar to that of 
the Dirac phenomenology optical potential shown by the dotted lines. 
In the peripheral region, furthermore, $R$ dependence of 
the two phenomenological optical potentials is similar to that of 
the microscopic one shown by the solid lines. 
At small $R$, the deviations of the microscopic optical potential from 
the phenomenological ones are small for $V_{\rm CE}$, but become 
sizable for $V_{\rm LS}$ and $W_{\rm CE}$. 
For $V_{\rm LS}$ the deviations little affect 
$d\sigma/d\Omega$ and $A_y$ up to 90$^{\circ}$ where 
the experimental data is available, but 
for $W_{\rm CE}$ the deviations slightly enhance 
$\sigma_{\rm R}$ and consequently overestimate the measured 
$\sigma_{\rm R}$ by several percentages.

\begin{figure}[htbp]
\begin{center}
 \includegraphics[width=0.35\textwidth,clip]{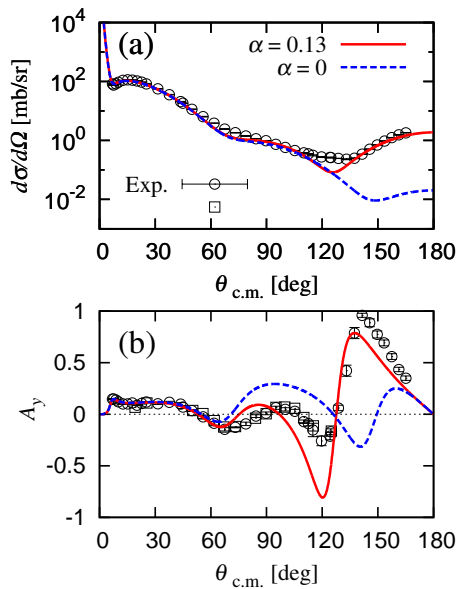}
 \caption{(Color online) 
 Effects of the parity-dependent potential 
 on (a) the elastic cross section $d\sigma/d\Omega$ 
 and (b) the vector analyzing power $A_y$ 
 for $p$+$^{4}$He scattering at $E=65$~MeV. 
 The solid (dashed) lines stand for the results of 
the $g$-matrix folding model with $\a=0.13$ ($\a=0$), where 
the phenomenological density is taken for $^{4}$He. 
Experimental data are taken from Refs.~\cite{pHe4-exp1,pHe4-exp2}.
    }
 \label{Fig-Xsec-4He}
\end{center}
\end{figure}

For the $p$+$^{4}$He system, 
the $g$-matrix folding model does not reproduce the measured elastic cross 
sections at backward angles $\theta_{\rm c.m.} \ga 90^{\circ}$. 
A possible origin of this deviation is the heavy-ion ($^{3}$H) exchange 
process that is not included in the folding model. 
The process generates a parity-dependent potential \cite{Votta:1974}.
We then rewrite $V_{\rm CE}$ as  
\bea
V_{\rm CE} \to V_{\rm CE}(1+\a (-1)^{L})
\eea
to incorporate the effect in the folding model phenomenologically. 
The parameter $\a$ is fitted to the experimental 
data~\cite{pHe4-exp1,pHe4-exp2}, and the resulting 
value is $\a=0.13$. 
In Fig.~\ref{Fig-Xsec-4He}, the solid (dashed) lines 
show $d\sigma/d\Omega$ and $A_y$ calculated 
by the $g$-matrix folding model with $\a=0.13$ ($\a=0$). 
Effects of the parity-dependent potential are appreciable 
at intermediate angles $60^{\circ} \la \theta_{\rm c.m.} \la 90^{\circ}$ and 
significant at backward angles at $\theta_{\rm c.m.} \ga 90^{\circ}$.

For the $p$+$^{12}$C system, the $g$-matrix folding model does not reproduce
the measured $A_y$ around $\theta_{\rm c.m.}=100^{\circ}$. 
Even the phenomenological optical 
potentials~\cite{Koning-Delaroche,Dirac1,Dirac2} cannot reproduce it. 
The parity-dependent potential does not solve this problem, since 
the potential enhances the elastic cross section largely at 
$\theta_{\rm c.m.} > 90^{\circ}$. Therefore this is an open problem. 
Except for the measured $A_y$ around $\theta_{\rm c.m.}=100^{\circ}$
for the $p$+$^{12}$C system, the $g$-matrix folding model yields 
good agreement with the data for targets of $A=4$--$208$.

Now we briefly analyze proton scattering at $E = 200$~MeV as 
an example of intermediate energies and a typical case of beam 
energies in Radioactive Isotope Beam Factory (RIBF).

\begin{figure}[htbp]
\begin{center}
 \includegraphics[width=0.4\textwidth,clip]{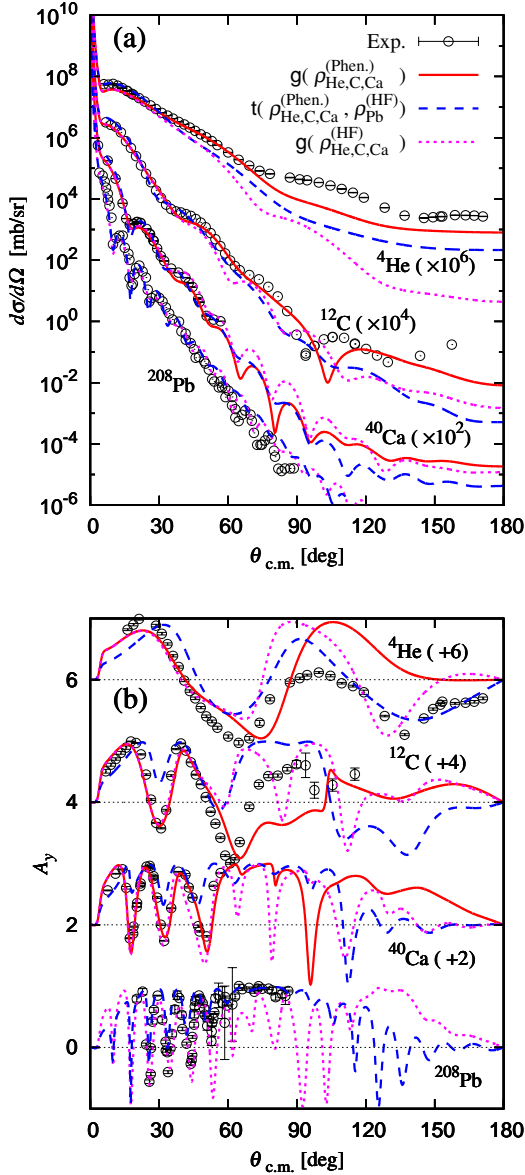}
 \caption{(Color online) Angular distributions of 
 (a) the elastic cross sections $d\sigma/d\Omega$ 
 and (b) the vector analyzing powers $A_y$ for proton scattering 
at $E = 200$~MeV from 
 $^{4}$He, $^{12}$C, $^{40}$Ca, and $^{208}$Pb targets. 
 See Fig. \ref{Fig-Xsec-stable} for the definition of lines.
 Experimental data are taken from Refs.~\cite{pHe4-200-exp1,pC12-200-exp1,pCa40-200-exp1,pPb208-200-exp1}.
 }
 \label{Fig-Xsec-stable-200}
\end{center}
\end{figure}

Figure~\ref{Fig-Xsec-stable-200} shows angular distributions 
of $d\sigma/d\Omega$ and $A_y$ for proton scattering from 
stable nuclei $^{4}$He, $^{12}$C, $^{40}$Ca, and $^{208}$Pb. 
The definition of lines is the same as in Fig. \ref{Fig-Xsec-stable}. 
The $g$-matrix folding model with the phenomenological density, 
shown by the solid lines, reproduces 
the data~\cite{pHe4-200-exp1,pC12-200-exp1,pCa40-200-exp1,pPb208-200-exp1} 
particularly at forward angles $\theta_{\rm c.m.} \la 60^{\circ}$. 
If Fig. \ref{Fig-Xsec-stable-200} is compared with Fig. \ref{Fig-Xsec-stable} as a function of the transferred momentum, 
$\theta_{\rm c.m.}=60^{\circ}$ in Fig. \ref{Fig-Xsec-stable-200}
corresponds to $\theta_{\rm c.m.}=128^{\circ}$ in Fig. \ref{Fig-Xsec-stable}. 
Therefore, the angle region $\theta_{\rm c.m.} > 60^{\circ}$ 
in Fig. \ref{Fig-Xsec-stable-200} corresponds to very backward angles 
$\theta_{\rm c.m.} > 128^{\circ}$ in Fig. \ref{Fig-Xsec-stable}. 
In this sense, it is natural that in Fig. \ref{Fig-Xsec-stable-200} 
the solid lines are deviated 
from the data at $\theta_{\rm c.m.} \ga 60^{\circ}$ particularly 
for lighter targets of $A \le 12$. 
Even at $\theta_{\rm c.m.} \ga 60^{\circ}$ 
for lighter targets of $A \le 12$, 
the $g$-matrix folding model with the phenomenological density, 
shown by the solid lines, 
yields better agreement with the data than that with 
the HF density denoted by the dotted lines. 
The medium effect, shown by the difference between the solid and dashed lines, 
is still significant at $E=200$~MeV.

\subsection{$p$+$^{6}$He scattering at 71 and 200~MeV}

\begin{figure}[htbp]
\begin{center}
 \includegraphics[width=0.35\textwidth,clip]{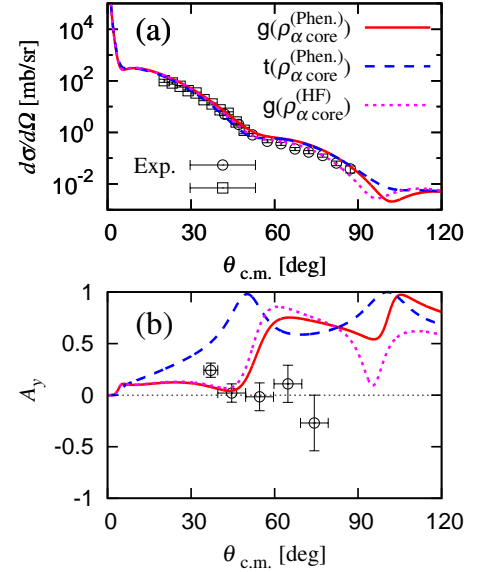}
 \caption{(Color online) 
 Angular distributions of (a) the elastic cross section 
 $d\sigma/d\Omega$ and (b) the vector analyzing power $A_y$ 
 for $p$+$^{6}$He scattering at $E=71$~MeV 
 calculated by the folding model with the $^4$He+$n$+$n$ model density. 
 The solid (dotted) lines represent the results 
 of the $g$-matrix folding model with 
 the phenomenological (HF) $^4$He density, whereas 
 the dashed lines correspond to the results of 
 the $t$-matrix folding model with  the phenomenological $^4$He density.
 Experimental data are taken from Ref.~\cite{Uesaka:2010,pHe6-exp1}.  
 }
 \label{Fig-Xsec-6He}
\end{center}
\end{figure}

As a typical example of proton scattering from unstable targets, we first 
consider $p$+$^{6}$He scattering at 71~MeV, since 
the experimental data are available not only for $d\sigma/d\Omega$ but also 
for $A_y$. The $^{6}$He density is obtained by 
the $^{4}$He+$n$+$n$ orthogonality-condition model; 
see Refs. \cite{Matsumoto3,Matsumoto4} 
for the details. As a $^{4}$He density in the model, 
we take either the phenomenological or the HF density. 
In Fig.~\ref{Fig-Xsec-6He}, the solid (dashed) lines show 
$d\sigma/d\Omega$ and $A_y$ calculated by 
the $g$-matrix ($t$-matrix) folding model with 
the phenomenological $^{4}$He density, whereas the dotted lines 
correspond to the results of the $g$-matrix folding model with 
the HF $^{4}$He density. 
The $g$-matrix folding model shown by the solid lines  
yields better agreement with the data~\cite{Uesaka:2010,pHe6-exp1} than 
the $t$-matrix folding model denoted by the dashed lines. 
The solid lines are consistent with the results of 
the nonlocal microscopic optical potential derived from 
the Melbourne $g$-matrix interaction~\cite{Uesaka:2010}, 
although $^{6}$He wave functions used in two calculations differ.

In Fig.~\ref{Fig-Xsec-6He}, the differences 
between the solid  and dotted lines come from 
the difference between the phenomenological and HF $^{4}$He densities. 
Comparing Fig.~\ref{Fig-Xsec-6He} with Fig.~\ref{Fig-Xsec-stable}, 
one can see that 
the difference in the $^{4}$He density yields smaller effects on 
$d\sigma/d\Omega$ and $A_y$ in $p$+$^{6}$He scattering 
than in $p$+$^{4}$He scattering. 
The difference in the $^{4}$He density is thus partly masked by 
densities of extra two neutrons in the $p$+$^{6}$He scattering.

\begin{figure}[htbp]
\begin{center}
 \includegraphics[width=0.5\textwidth,clip]{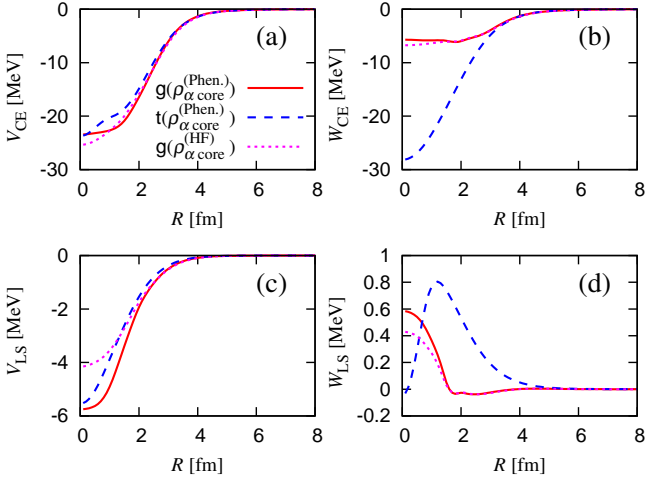}
 \caption{(Color online) $R$ dependence of the microscopic optical potential 
 for $p$+$^{6}$He scattering at $E=71$~MeV. 
 See Fig. \ref{Fig-Xsec-6He} for the definition of lines.
 Panels (a), (b), (c), and (d) correspond to $V_{\rm CE}$,
 $W_{\rm CE}$, $V_{\rm LS}$, and $W_{\rm LS}$, respectively. 
    }
 \label{Fig-U-6He}
\end{center}
\end{figure}

In Fig.~\ref{Fig-Xsec-6He}, the nuclear-medium effect is 
shown by the differences between the solid and dashed lines, and 
the effect is much more significant for $A_y$ than $d\sigma/d\Omega$. 
The differences mainly come from the fact that 
$W_{\rm CE}$ calculated with the $g$-matrix folding model is much 
less absorptive than 
that with the $t$-matrix folding model; see Fig. \ref{Fig-U-6He} 
for $R$ dependence of $U$.

As mentioned above, the $g$-matrix folding model yields 
better agreement with the data than the $t$-matrix folding model. 
The agreement is, however, not perfect particularly for $A_y$. 
The parity-dependent potential little works at forward angles 
$\theta < 90^{\circ}$ in which the experimental data are available. 
We then make a potential search 
to see what causes the difference between the theoretical and experimental 
results. Here we change the strength of $U$ as 
\bea
&&U \to (f_V^{\rm CE} V_{\rm CE} +i f_W^{\rm CE} W_{\rm CE}) \nonumber \\
&&~~~~~~+ (f_{V}^{\rm LS} V_{\rm LS}+if_{W}^{\rm LS} W_{\rm LS}) 
{\bfi L} \cdot {\bfi \sigma} + V_{\rm Coul} 
\eea
with adjustable parameters 
$f_V^{\rm CE}, f_W^{\rm CE}, f_{V}^{\rm LS}, f_{W}^{\rm LS}$. 
A best fit is obtained at 
$(f_V^{\rm CE}, f_W^{\rm CE}, f_{V}^{\rm LS}, f_{W}^{\rm LS})
=(0.7,0.7,0.9,4.0)$, 
but $d\sigma/d\Omega$ and $A_y$ are not so sensitive at $\theta > 40^{\circ}$ 
to variations of $f_W^{\rm CE}$ and $f_{V}^{\rm LS}$ and hence 
these parameters are not determined sharply.

\begin{figure}[htbp]
\begin{center}
 \includegraphics[width=0.35\textwidth,clip]{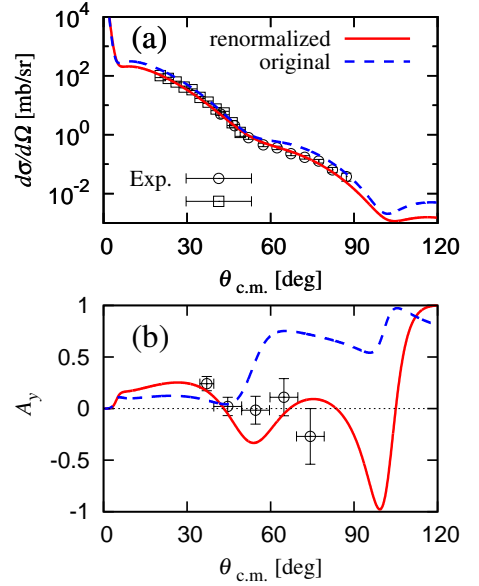}
 \caption{(Color online) 
 Potential search for $p$+$^{6}$He scattering at $E=71$~MeV 
 in (a) the elastic cross sections $d\sigma/d\Omega$ and (b) the vector analyzing powers $A_y$.    
 The solid lines represent the best-fit result with 
 $(f_V^{\rm CE}, f_W^{\rm CE}, f_{V}^{\rm LS}, f_{W}^{\rm LS})
=(0.7,0.7,0.9,4.0)$, 
 whereas the dashed lines correspond to the original results. 
 Experimental data are taken from Refs.~\cite{Uesaka:2010,pHe6-exp1}.  
}
 \label{Fig-6He-U-search}
\end{center}
\end{figure}

Figure \ref{Fig-6He-U-search} shows the potential search for 
$p$+$^6$He scattering at $E$ = 71 MeV. 
The solid lines represent the results of the potential search, 
whereas the dashed lines mean the results of the $g$-matrix folding model with the phenomenological $^4$He density. 
A small reduction of $d\sigma/d\Omega$ from 
the dashed to the solid line at $\theta > 50^{\circ}$ 
mainly comes from the reduction of $V_{\rm CE}$, 
and a large change of $A_y$ from the dashed to the solid line 
at $\theta > 40^{\circ}$ 
is mainly originated in the increase of $W_{\rm LS}$. Thus 
the large enhancement of $W_{\rm LS}$ is necessary, but it is not clear 
what causes the large enhancement of $W_{\rm LS}$. 
This is an interesting question to be solved in the future.

\begin{figure}[htbp]
\begin{center}
 \includegraphics[width=0.35\textwidth,clip]{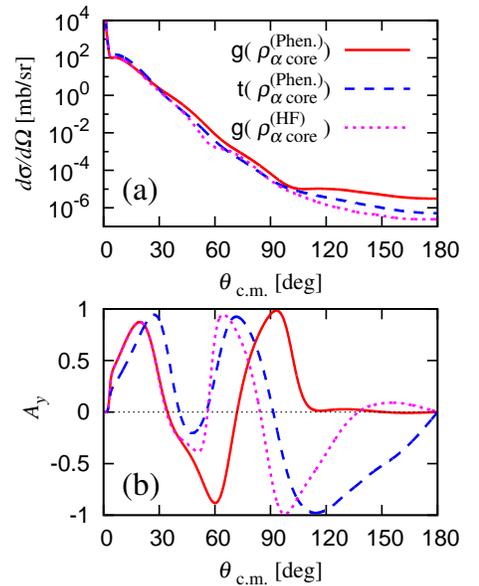}
 \caption{(Color online) Angular distributions of   
 (a) the elastic cross section $d\sigma/d\Omega$ 
 and (b) the vector analyzing power $A_y$  
 for proton scattering from $^{6}$He at $E = 200$~MeV. 
 See Fig. \ref{Fig-Xsec-6He} for the definition of lines.
   }
 \label{Fig-Xsec-6He-200}
\end{center}
\end{figure}

Now we briefly analyze proton scattering from $^{6}$He 
at $E = 200$~MeV. Any experimental data is not available for this energy 
at the present stage, but the measurements particularly of $A_y$ are 
planned in RIBF. 
Figure~\ref{Fig-Xsec-6He-200} shows angular distributions 
of $d\sigma/d\Omega$ and $A_y$ for proton scattering from $^{6}$He 
at $E = 200$~MeV. 
The definition of line is the same as in Fig. \ref{Fig-Xsec-6He}. 
As shown in Fig. \ref{Fig-Xsec-stable-200}, 
the $g$-matrix folding model shown by the solid lines yields consistent results with the data for stable target nuclei. 
Hence, the model prediction, shown by the solid lines 
in Fig. \ref{Fig-Xsec-6He-200}, is expected to be reliable also 
for $^{6}$He, particularly 
at forward angles $\theta_{\rm c.m.} \la 60^{\circ}$. 
At backward angles $\theta_{\rm c.m.} \ga 60^{\circ}$, 
the solid lines differ from the dashed and dotted 
lines. This means that it is necessary to include the medium effect and 
use the phenomenological $^{4}$He density there.

\subsection{$A$ dependence of microscopic optical potentials}
\label{Global properties}

The $g$-matrix folding model is successful in reproducing the data on 
polarized proton scattering for targets of $A=12$--$208$. 
In general, the $g$-matrix folding model is considered to be 
less accurate for light targets 
such as $^{4,6}$He, since the $g$-matrix interaction is evaluated in nuclear matter and 
the local density approximation used seems to be more reliable 
for heavier targets. 
Nevertheless, the agreement of the model calculations with the data 
is still reasonable for light targets $^{4,6}$He. 
We then derive global properties of the local microscopic optical potentials 
for proton scattering at 65~MeV. The potentials are calculated with 
the $g$ matrix folding model with the phenomenological target densities 
for $A \le 40$ and the HF densities for $A>40$.

Since the microscopic optical potentials do not have any simple shape, 
we consider the volume integral $J$ and the rms radius ${\langle R^2 \rangle}^{\frac{1}{2}}$ for each part of $U$: 
\bea
J=\int  F(R) d \vR,~~~ {\big\langle R^2 \big\rangle}^{\frac{1}{2}}=\sqrt{\int R^2 F(R)d \vR /J},
\eea
where $F(R)$ represents each part of $U$. 
As shown in Figs. \ref{Fig-Pot-40Ca} and \ref{Fig-U-6He}, 
$W_{\rm LS}$ is much smaller than $V_{\rm LS}$ and hence yields no 
significant effect on the present systems. Furthermore, 
meanings of the volume integral and the rms radius are not clear 
for $W_{\rm LS}$, since $W_{\rm LS}$ is an oscillating function of $R$. 
For these reasons, we do not consider the volume integral and the rms radius 
of $W_{\rm LS}$ here.

\begin{figure}[htbp]
\begin{center}
 \includegraphics[width=0.4\textwidth,clip]{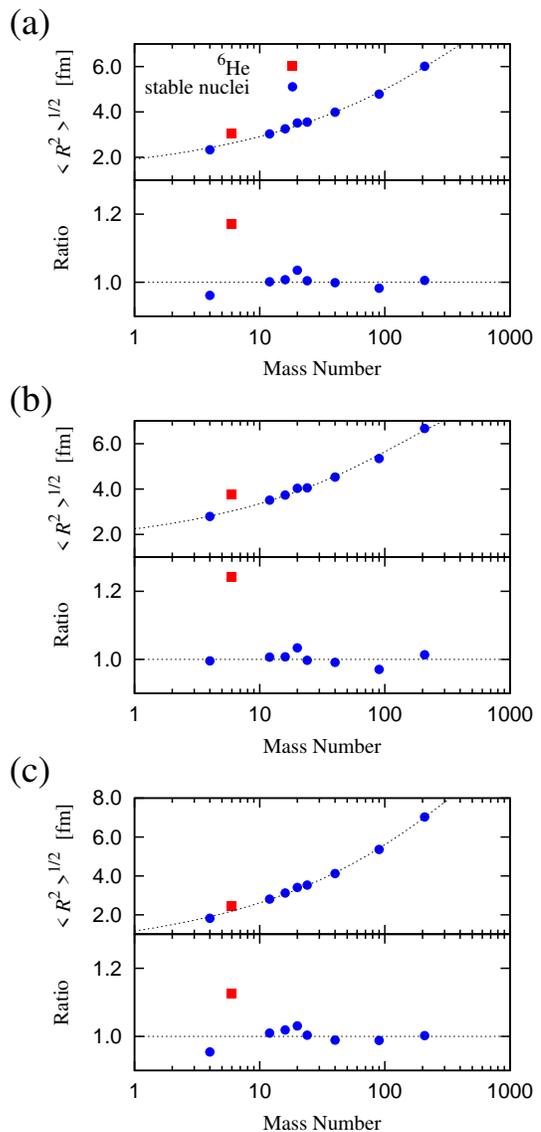}
 \caption{(Color online) 
 $A$ dependence of the rms radius of the microscopic 
 optical potential at 65~MeV.  
 The $A$ dependence is shown by the circles for stable targets and by the squares for $^6$He target. 
 Panels (a), (b), and (c) correspond to 
 $V_{\rm CE}$, $W_{\rm CE}$, and $V_{\rm LS}$, respectively. 
 See Table \ref{tab:Parameters fitting for the RMS radius of U} 
 for the parameter set of the stable-nucleus (dotted) line.
  }
 \label{Fig-U-RMS}
\end{center}
\end{figure}

Figure~\ref{Fig-U-RMS} shows the rms radius of $U$ 
as a function of $A$. 
For each part of $U$, $A$ dependence of the rms radius is determined 
for stable targets of $A=4$--$208$ by a curve 
\bea
a_1 A + a_2 A^{2/3}  + a_3 A^{1/3} + a_4 ;
\eea
see Table \ref{tab:Parameters fitting for the RMS radius of U}
for the results of the fitting. 
The fitting line is referred to as the stable-nucleus line. 
For each panel, the lower figure shows a ratio of the rms radius to 
the stable-nucleus line. Since $^6$He has a halo structure, 
the $^6$He rms radius is enlarged by the structure 
from the stable-nucleus line 
by $\sim$18\% for $V_{\rm CE}$, 
$\sim$25\% for $W_{\rm CE}$ and 
$\sim$12\% for $V_{\rm LS}$. 
The large enhancement of $^6$He rms radius in $W_{\rm CE}$ 
means that the reaction cross section is a good tool of 
searching for halo nuclei and determining their radii.

\begin{table}
\begin{center}
\begin{tabular}{|c|c|c|c|c|}
 \hline
 &$a_1$ &$a_2$ &$a_3$ &$a_4$ \\
 \hline
 $V_{CE}$&-0.0023 &0.0104 &0.8508 &1.0578 \\
 $W_{CE}$&-0.0076 &0.0473 &0.8732 &1.332 \\
 $V_{LS}$&-0.0033 &0.0137 &1.2305 &-0.0682 \\
 \hline
\end{tabular}
\end{center}
\caption{Parameter fitting for $A$ dependence of the rms radius of $U$ 
with a functional form $a_1 A + a_2 A^{2/3}  + a_3 A^{1/3} + a_4$. 
All the parameters are presented in units of fm. 
}
\label{tab:Parameters fitting for the RMS radius of U}
\end{table}

\begin{figure}[htbp]
\begin{center}
 \includegraphics[width=0.4\textwidth,clip]{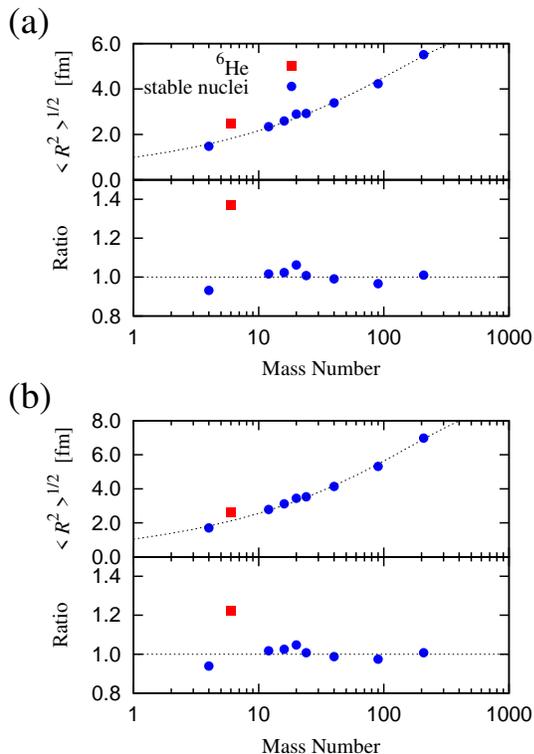}
 \caption{(Color online) 
 $A$ dependence of rms radii of (a) $\rho$ 
 and (b) $d\rho/dr$. 
 See Fig. \ref{Fig-U-RMS} for the definition of symbols  
 and Table \ref{tab:Parameters fitting for rho} for 
the parameter set of the stable-nucleus (dotted) line. 
   }
 \label{Fig-rho-RMS}
\end{center}
\end{figure}

We consider the reason why the enhancement of $^6$He rms radius is larger 
in $V_{\rm CE}$ than in $V_{\rm LS}$. For this purpose, 
we show $A$ dependence of rms radii 
of $\rho$ and $d \rho/d r$ in Fig. \ref{Fig-rho-RMS}. 
In the short-range limit of the $g$-matrix interaction, 
the shape of $V_{\rm CE}$ 
($V_{\rm LS}$) is approximately determined by that of $\rho$ ($d \rho/d r$). 
The enhancement of $^6$He rms radius 
from the stable-nucleus line is smaller for $d \rho/d r$ 
than for $\rho$, as shown in Fig. \ref{Fig-rho-RMS}; 
see Table \ref{tab:Parameters fitting for rho} for 
the stable-nucleus (dotted) line.

\begin{table}
\begin{center}
\begin{tabular}{|c|c|c|c|c|}
 \hline
 &$a_1$ &$a_2$ &$a_3$ &$a_4$ \\
 \hline
 $\rho$&-0.0074 &0.0378 &0.9578 &-0.0022 \\
 $d\rho/dr$&-0.0079 &0.0402 &1.2478 &-0.2322 \\
 \hline
\end{tabular}
\end{center}
\caption{Parameter fitting for $A$ dependence of rms radii of 
$\rho$ and $d\rho/dr$ with a functional form 
$a_1 A + a_2 A^{2/3}  + a_3 A^{1/3} + a_4$. 
All the parameters are presented in units of fm. 
}
\label{tab:Parameters fitting for rho}
\end{table}

\begin{figure}[htbp]
\begin{center}
 \includegraphics[width=0.4\textwidth,clip]{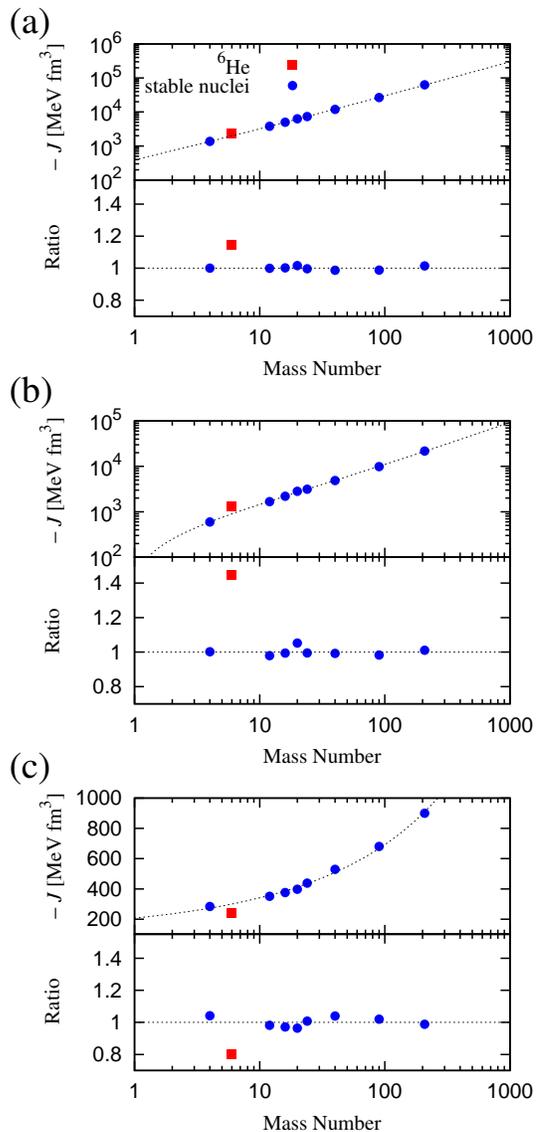}
 \caption{(Color online) 
 $A$ dependence of the volume integral of the microscopic 
 optical potential at 65~MeV. 
 Panels (a), (b), and (c) correspond to 
 $V_{\rm CE}$, $W_{\rm CE}$, and $V_{\rm LS}$, respectively.
 See Fig. \ref{Fig-U-RMS} for the definition of symbols
 and Table \ref{tab:Parameters fitting for volume integral} 
 for the parameter set of the stable-nucleus (dotted) line.
  }
 \label{Fig-U-Vol}
\end{center}
\end{figure}

Figure~\ref{Fig-U-Vol} shows 
the volume integral of $U$ as a function of $A$, 
and the parameters of the fits for the stable nuclei are shown 
in Table~\ref{tab:Parameters fitting for volume integral}. 
The volume integral is also enhanced at $A=6$ 
from the stable nucleus line by the halo structure of $^6$He.
This enhancement is larger in $W_{\rm CE}$ than in $V_{\rm CE}$. 
In $V_{\rm LS}$, meanwhile, the volume integral is suppressed at 
$A=6$ from the stable nucleus line. The halo structure of $^6$He makes 
$\rho$ broader and, hence, suppresses $d\rho/dr$ and, consequently, 
weakens the strength of $V_{\rm LS}$.

\begin{table}
\begin{center}
\begin{tabular}{|c|c|c|c|c|}
 \hline
 &$a_1$ &$a_2$ &$a_3$ &$a_4$ \\
 \hline
 $V_{CE}$&301.28 &-59.695 &272.22 &-125.78 \\
 $W_{CE}$&81.082 &92.201 &301.22 &-439.22 \\
 $V_{LS}$&0.093 &6.24 &95.645 &105.11 \\
 \hline
\end{tabular}
\end{center}
\caption{Parameter fitting for $A$ dependence of the volume integral of $U$ 
with a functional form $a_1 A + a_2 A^{2/3}  + a_3 A^{1/3} + a_4$. 
All the parameters are presented in units of MeV fm$^3$. 
}
\label{tab:Parameters fitting for volume integral}
\end{table}

\begin{figure}[htbp]
\begin{center}
\includegraphics[width=0.4\textwidth,clip]{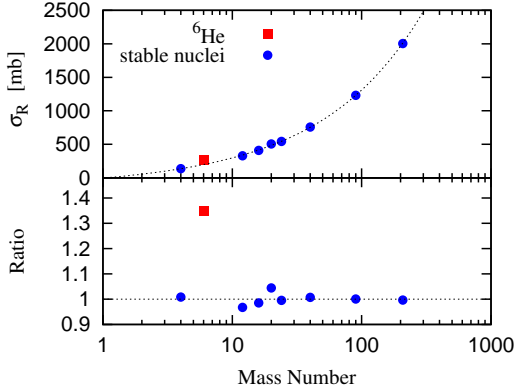}
 \caption{(Color online) 
 $A$ dependence of the reaction cross sections 
 for proton scattering  at $E = 65$~MeV. 
  See Fig. \ref{Fig-U-RMS} for the definition of symbols
 and Table \ref{tab:Parameters fitting for sigma}
 for the parameter set of the stable-nucleus (dotted) line.
    }
 \label{Fig-sigma-R}
\end{center}
\end{figure}

Finally, we show $A$ dependence of $\sigma_{\rm R}$ in 
Fig.~\ref{Fig-sigma-R} for proton scattering at $E = 65$~MeV. 
The reaction cross section is enhanced for $^6$He by $\sim 35$\% 
from the stable-nucleus (dotted) line; 
see Table \ref{tab:Parameters fitting for sigma} for the parameter set. 
This enhancement is a result of the corresponding enhancement of $W_{\rm CE}$ 
in its rms radius and volume integral.

\begin{table}
\begin{center}
\begin{tabular}{|c|c|c|c|c|}
\hline
&$a_1$ &$a_2$ &$a_3$ &$a_4$ \\
\hline
$\sigma_{\rm R}$&-0.6797 &45.534 &122.34 &-170.78 \\
\hline
\end{tabular}
\end{center}
\caption{Parameter fitting for $A$ dependence of $\sigma_{\rm R}$ 
with a functional form $a_1 A + a_2 A^{2/3}  + a_3 A^{1/3} + a_4$. 
All the parameters are presented in units of mb. 
}
\label{tab:Parameters fitting for sigma}
\end{table}

Throughout all the analyses in this subsection, 
one can find that the broadening of $^{6}$He density 
due to halo structure (weak-binding nature) 
yields significant effects on $U$. 
The effect differs between the central and spin-orbit parts of $U$. 
For the central part, it enhances the strength and 
the rms radius by about $20\% \sim 40\%$. 
For $V_{\rm LS}$, meanwhile, the broadening 
enlarges the rms radius by about $10\%$ but weakens the strength 
by about $20\%$.

\subsection{Isotope dependence of microscopic optical potential}
\label{Ne isotopes}

The broadening of matter densities is induced by the weak-binding nature 
for unstable nuclei and by the deformation for largely deformed nuclei.
Deformation is a picture in the body-fixed frame. Hence, 
the deformed density in the body-fixed frame should be transformed 
into the corresponding ground-state density 
in the space-fixed frame by making the angular-momentum projection. 
This was done properly in Eq. \eqref{eq:amddens}. 
The projection makes the spherical ($\lambda=0$) part 
of the ground-state density broad.

The local-potential version of the Melbourne $g$-matrix folding model is 
successful in reproducing both $^{20-32}$Ne+$^{12}$C scattering and 
proton scattering from stable nuclei at 65 MeV. This indicates that 
the model is reliable also 
for proton scattering from $^{20-32}$Ne at 65 MeV. 
Now we analyze the proton scattering with the local-potential version 
of the folding model in order to investigate how large weak-binding and 
deformation effects are on $U$ and $\sigma_{\rm R}$. 
The analysis is particularly interesting for $^{31}$Ne, 
since it is a one-neutron halo nucleus with large deformation.

\begin{figure}[htbp]
\begin{center}
 \includegraphics[width=0.4\textwidth,clip]{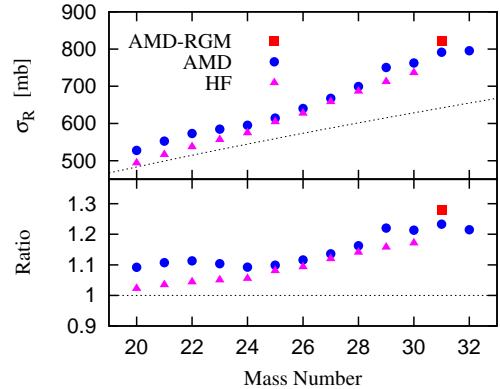}
 \caption{(Color online) 
 Isotope dependence of the reaction cross sections 
 for $p$+$^{20-32}$Ne scattering at 65 MeV.  
 The circles (triangles) show the results of 
 the AMD (HF) densities,
 and the square for $^{31}$Ne is the result of the AMD-RGM density.
 See Table \ref{tab:Parameters fitting for sigma} 
 for the parameter set of the stable-nucleus (dotted) line. 
  }
 \label{Fig-sigma-R-Ne}
\end{center}
\end{figure}

Figure~\ref{Fig-sigma-R-Ne} shows isotope dependence of $\sigma_{\rm R}$ for 
$p$+$^{20-32}$Ne scattering at 65 MeV. 
The dotted line denotes the stable-nuclei line for $\sigma_{\rm R}$; 
see Table \ref{tab:Parameters fitting for sigma}
for the results of the fitting. 
The triangles represent the results of the $g$-matrix folding model with 
the spherical HF densities, whereas 
the closed circles correspond to those with the AMD densities in which 
Ne isotopes are deformed. 
The deviation of the triangles from the dotted line represents 
the weak-binding effect, whereas 
that of the closed circles from the triangles shows 
the deformation effect. 
Here, the weak-binding effect includes the effects of neutron skin. 
These effects are more apparent in the lower panel where 
the $\sigma_{\rm R}$ for $^{A}$Ne are normalized by those 
for stable nuclei with the same $A$. 
At $20 \le A \le 23$, the deformation effect is more significant 
than the weak-binding effect. 
Since the weak-binding effect becomes large as $A$ increases, 
the effect is more important than the deformation effect at 
$24 \le A \le 30$. 
For $A=31, 32$, Ne isotopes have no bound state 
in the spherical HF calculation. 
In this sense, both the deformation and the weak-binding effect 
are important there.

For $^{31}$Ne, the one-neutron separation energy is quite small. 
We then do a AMD-RGM calculation to make a 
tail correction to the AMD density. 
The closed square denotes the result of the AMD-RGM density, and 
the deviation of the closed square from the stable-nucleus line 
represents a net effect of halo structure and deformation. 
This effect is detectable from the measurements of 
$\sigma_{\rm R}$ or the one-neutron removal cross section. 
Actually, this was done for the $^{31}$Ne + $^{12}$C system 
at 240 MeV/nucleon \cite{Nakamura,Takechi-Ne,Min11,Min12,Sum12}.

\begin{figure}[htbp]
\begin{center}
 \includegraphics[width=0.4\textwidth,clip]{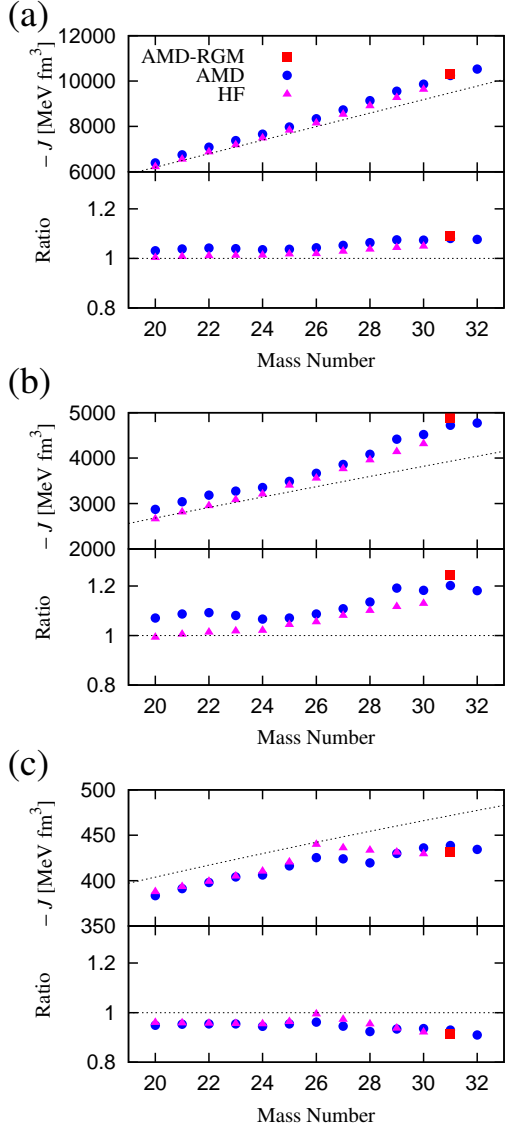}
 \caption{(Color online) 
  $A$ dependence of the volume integral of the microscopic 
 optical potential for $p$+$^{20-32}$Ne scattering at 65~MeV.
 Panels (a), (b) and (c) correspond to $V_{\rm CE}$, $W_{\rm CE}$
 and $V_{\rm LS}$, respectively.
 See Fig. \ref{Fig-sigma-R-Ne} for the definition of symbols 
 and Table \ref{tab:Parameters fitting for volume integral} 
 for the parameter set of the stable-nucleus (dotted) line. 
 }
 \label{Fig-U-Ne-Vol}
\end{center}
\end{figure}

\begin{figure}[htbp]
\begin{center}
 \includegraphics[width=0.4\textwidth,clip]{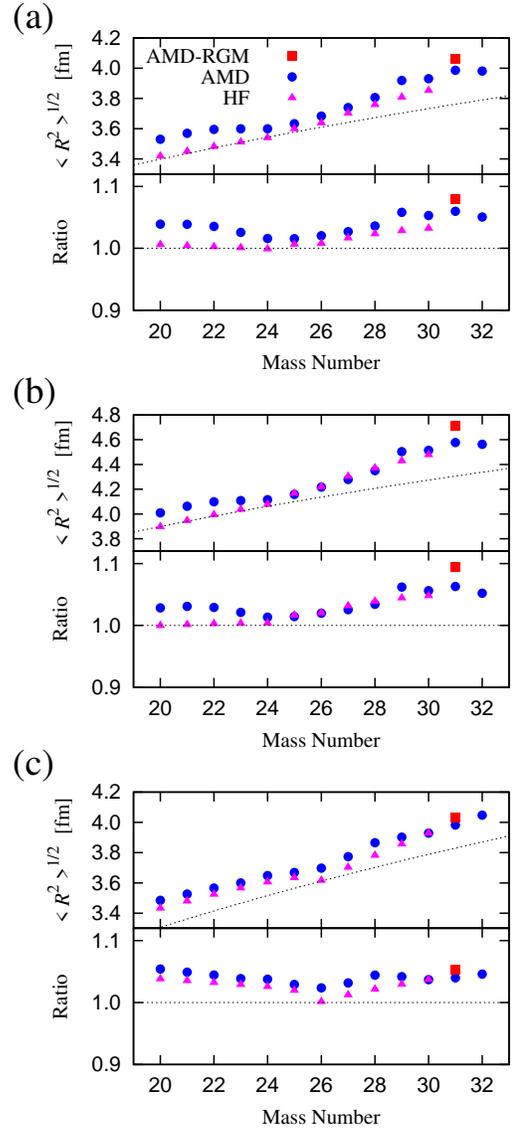}
 \caption{(Color online) 
 $A$ dependence of the rms radius of the microscopic 
 optical potential for $p$+$^{20-32}$Ne scattering at 65~MeV. 
  Panels (a), (b) and (c) correspond to 
 $V_{\rm CE}$, $W_{\rm CE}$ and $V_{\rm LS}$, respectively.
 See Fig. \ref{Fig-sigma-R-Ne} for the definition of symbols
 and Table \ref{tab:Parameters fitting for the RMS radius of U} 
 for the parameter set of the stable-nucleus (dotted) line.
  }
 \label{Fig-U-Ne-RMS}
\end{center}
\end{figure}

Figures~\ref{Fig-U-Ne-Vol} and \ref{Fig-U-Ne-RMS} show the volume integral 
and the rms radius of $U$ 
for $p$+$^{20-32}$Ne scattering at 65~MeV, respectively. 
The weak-binding and nuclear-deformation effects play the same role between 
$U$ and $\sigma_{\rm R}$. 
Actually, the deformation effect is more important than 
the weak-binding effect at $20 \le A \le 23$, whereas the latter is more 
significant than the former at $24 \le A \le 30$. 
The density-broadening effect due to weak-binding nature and deformation 
in Ne isotopes is the same as that due to weak-binding nature in $^{6}$He. 
For $^{31}$Ne as a typical case, it suppresses the volume integral by 
about $10\%$ and enlarges the rms radius by about $5\%$ for $V_{\rm LS}$, 
whereas it enlarges both the volume integral and the rms radius 
by about $10\%$ for $V_{\rm CE}$ and $W_{\rm CE}$. 
The imaginary central part $W_{\rm CE}$ is more sensitive to 
the density-broadening effect than the other parts. Thus 
$\sigma_{\rm R}$ is a good quantity to investigate the density broadening.

\subsection{Nuclear deformation effects on elastic scattering}
\label{Deformed}

If a target nucleus is strongly deformed, the deformation affects 
proton elastic scattering. For example, 
$^{22,30}$Ne and their neighborhood nuclei are known to be well deformed. 
For even target nuclei with $I=0$, nuclear deformation yields 
two kinds of effects. 
First, it enlarges the spherical part of the target density 
and then the spherical part of $U$, as mentioned in Sec. \ref{Ne isotopes}.   
Second, the deformation enhances target excitations and then 
the back reactions to the elastic channel. 
The target-excitation effects are investigated in this subsection.

We consider proton elastic scattering from $^{22,30}$Ne, and 
rotational excitations to the $2^{+}$ excited states are treated 
with the coupled-channel method. 
Transition densities between the elastic and inelastic channels are 
constructed by the deformed Woods-Saxon (DWS) model~\cite{Sum12} 
with the quadrupole 
deformation parameter ($\beta_2^{}$) evaluated by AMD; here 
the other parameters of the DWS model are already determined to 
reproduce spectroscopic properties of high-spin states from light 
to heavy deformed nuclei, e.g., the quadrupole moments, 
the moments of inertia and the rms radii.
Coupling potentials between the elastic and inelastic channels
are obtained by folding the Melbourne $g$-matrix interaction
with the transition densities.

\begin{figure}[htbp]
\begin{center}
 \includegraphics[width=0.4\textwidth,clip]{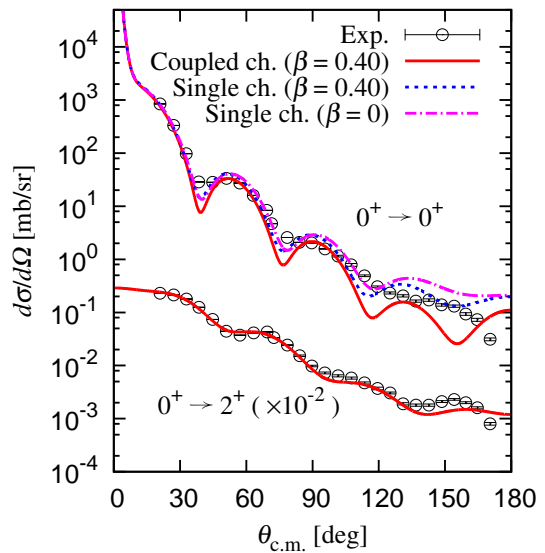}
 \caption{(Color online) 
 Angular distributions of the differential cross sections for proton elastic and inelastic scattering from 
$^{22}$Ne at 35~MeV. The solid lines are the results of the coupled-channel 
calculations, while the dotted and dot-dashed lines correspond to the results 
of the single-channel calculations with and without deformation. 
The experimental data are taken from Ref.~\cite{p-22Ne-inelastic}. 
  }
 \label{dcs-pNe22_2+}
\end{center}
\end{figure}

Figure~\ref{dcs-pNe22_2+} shows angular distributions for $p$+$^{22}$Ne 
scattering at 35~MeV. 
In this case, AMD yields $\beta_2^{}=0.4$~\cite{Sum12}. 
The solid lines denote the results of the coupled-channel calculations, 
while the dotted and dot-dashed lines correspond to the 
results of the single-channel calculations with and without deformation, 
respectively. 
The solid line reproduces the experimental data~\cite{p-22Ne-inelastic} 
on both the elastic and the inelastic scattering with no adjustable parameter. 
The deviation of the solid line from the dotted line shows 
target-excitation effects, while the deviation of the dotted line 
from the dot-dashed one corresponds to density-broadening effects due to 
deformation. These effects are small at forward angles 
$\theta_{\rm c.m.} \la 60^{\circ}$, but become appreciable at 
middle and backward angles $\theta_{\rm c.m.} \ga 60^{\circ}$, although 
the target-excitation effects are relatively larger than 
the density-broadening effects. As for $\sigma_{\rm R}$, 
the target-excitation effects are 4~\%, whereas 
the density-broadening effects are about 5~\%. 
The two effects are thus comparable for $\sigma_{\rm R}$ 
at this lower incident energy.

\begin{figure}[htbp]
\begin{center}
 \includegraphics[width=0.4\textwidth,clip]{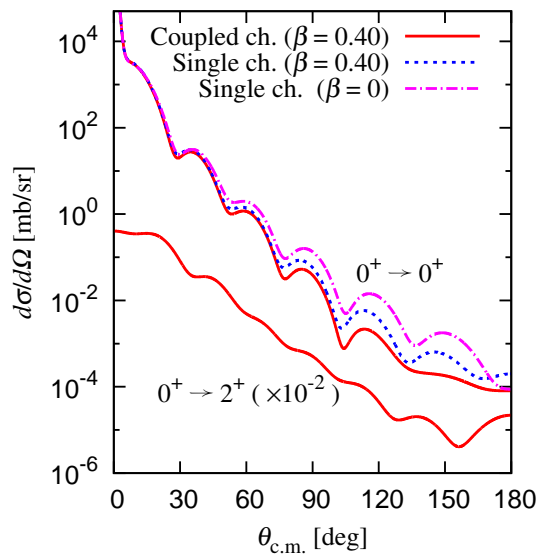}
 \caption{(Color online) 
Angular distributions of the differential cross sections for proton elastic and inelastic scattering from 
$^{30}$Ne at 65~MeV. 
See Fig. \ref{dcs-pNe22_2+} for the definition of lines. 
  }
 \label{dcs-pNe30_2+}
\end{center}
\end{figure}

Similar analyses are made in Fig.~\ref{dcs-pNe30_2+} 
for $p$+$^{30}$Ne scattering at 65~MeV; 
here AMD yields $\beta_2^{}=0.4$~\cite{Sum12}. 
Target-excitation and density-broadening effects become appreciable 
at intermediate and backward angles $\theta_{\rm c.m.} \ga 50^{\circ}$, 
the two effects are comparable there. 
As for $\sigma_{\rm R}$, the target-excitation effects are 2~\%, 
while the density-broadening effects are about 4~\%. 
At relatively higher incident energies such as 65~MeV, 
target-excitation effects are thus as small as 
density-broadening effects due to deformation, whereas 
density-broadening effects due to weak binding are about 20~\% 
as shown in Fig.~\ref{Fig-sigma-R-Ne}. 
Density-broadening effects due to weak binding 
thus become most significant for unstable target nuclei.

For odd target nuclei with $I > 0$, nuclear deformation induces 
another effect in addition to density-broadening and target-excitation 
effects. Namely the deformation makes the folding 
potential $U$ non spherical, as seen in Eq.~\eqref{eq:amddens}, 
and consequently induces a reorientation of target spin in 
proton scattering. We briefly investigate the reorientation effects here.

\begin{figure}[htbp]
\begin{center}
 \includegraphics[width=0.4\textwidth,clip]{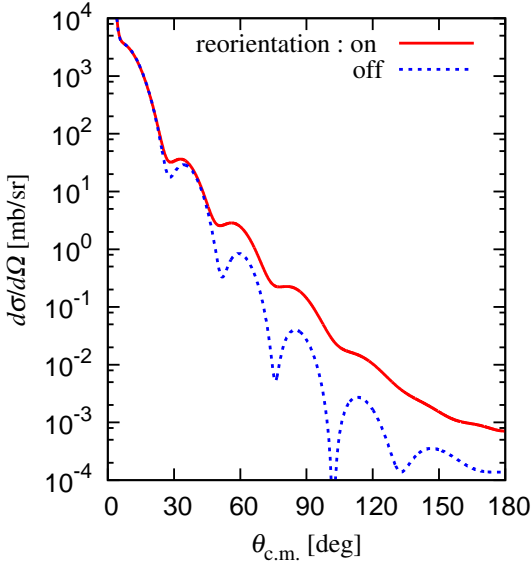}
 \caption{(Color online) 
 Angular distribution of the elastic cross section $d\sigma/d\Omega$ for $p$+$^{31}$Ne elastic scattering at 65 MeV. 
 The solid (dotted) line shows the result of the folding-model calculation 
 with (without) reorientation effects. 
  }
 \label{Fig-reorientation}
\end{center}
\end{figure}

For targets with $I={1/2}^{\pm}$, we numerically confirmed 
that the reorientation effects are quite small 
on both reaction cross sections and differential elastic cross sections. 
Then we consider the case of $I={3/2}^{-}$ here. 
Figure~\ref{Fig-reorientation} shows differential cross sections 
for proton elastic scattering from a $^{31}$Ne$({3/2}^{-})$ target at 65 MeV. 
Here the deformed target density, which
is calculated by the DWS model with $\beta_2^{}=0.42$ evaluated by AMD, 
is transformed into the corresponding ground-state 
density by the angular momentum projection. 
For simplicity, we neglect the spin-orbit interaction. 
The solid and dotted lines are the results of 
the folding-model calculations with and without reorientation effects, 
respectively. The effects are quite small at forward angles 
$\theta_{\rm c.m.} \la 50 ^{\circ}$ and become significant at middle and backward 
angles $\theta_{\rm c.m.} \ga 50 ^{\circ}$. Meanwhile, the effects are 
less than 1\% on reaction cross sections. 
Reorientation effects are thus important for differential cross sections 
at middle and backward angles, but quite small for reaction cross sections.

\subsection{Isovector components of optical potentials}
\label{Isovector}

In this subsection, we discuss effects of the isovector component of $U$ 
briefly. 
For simplicity, we neglect the spin-orbit part of $U$, since 
the effect is negligible for the $p$+$^{20-32}$Ne systems at 65 MeV 
analyzed here. 
In general, the central part of nucleon optical potentials 
can be decomposed into the isoscalar and isovector components, $U_0$ and $U_1$, 
as
\bea
U(\vR)=U_0(\vR) + 4 U_1(\vR)\frac{\vt\cdot\vT}{A},
\label{Eq:Lane-potential}
\eea
where $\vt$ and $\vT$ denote the projectile and target isospins, respectively. 
Proton and neutron optical potentials, $U_p$ and $U_n$, 
can be described by 
the isoscalar and isovector components as 
\bea
U_p&=&U_0-\frac{N-Z}{A}U_1, 
\label{Up-isovector}
\\
U_n&=&U_0+\frac{N-Z}{A}U_1, 
\eea
and hence 
\bea
U_0&=&\frac{U_p+U_n}{2}, \\
U_1&=&\frac{U_n-U_p}{2}\frac{A}{N-Z}, 
\label{eq:U1}
\eea
where 
$N$ and $Z$ are the neutron and proton numbers of target nucleus. 
In the folding model, the difference $U_n-U_p$ is proportional to 
the difference $\rho_n-\rho_p$ and hence the isovector component $U_1$ 
of Eq.~\eqref{eq:U1} is to 
the factor $(\rho_n-\rho_p)/(N-Z)$ that little depends on $N-Z$. 
Therefore it follows from Eq.~\eqref{Up-isovector} 
that the contribution of $U_1$ to $U_p$ linearly increases as $N-Z$ becomes 
large. In the folding model, the isoscalar part $U_0$ is proportional to 
the total density $\rho_n+\rho_p$. Hence the contribution of $U_0$ to $U_p$ 
also becomes large as $N$ increases with $Z$ fixed, since so does $\rho_n$.

\begin{figure}[htbp]
\begin{center}
 \includegraphics[width=0.4\textwidth,clip]{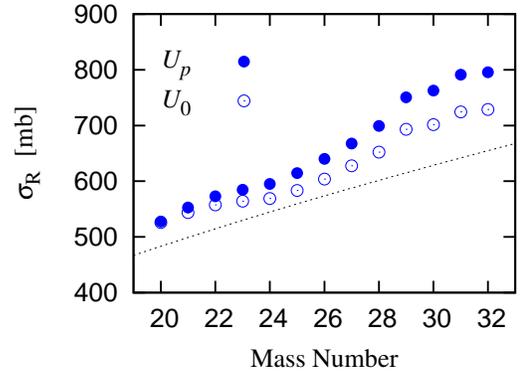}
 \caption{(Color online) 
Effects of the isoscalar and isovector components, $U_0$ and $U_1$, 
on reaction cross sections 
for proton scattering from $^{20-32}$Ne system at 65~MeV. 
The open (closed) circles denote the results 
of $U_0$ ($U_p$) calculated with the AMD densities. 
The dotted line means the stable-nucleus line. 
  }
 \label{ReacXsec-decomp}
\end{center}
\end{figure}

Figure~\ref{ReacXsec-decomp} shows reaction cross sections for 
$p$+$^{20-32}$Ne scattering at 65~MeV. 
The reaction cross sections are calculated with either $U_p$ or $U_0$. 
The closed (open) circles denote the results 
of $U_p$ ($U_0$) calculated with the AMD densities. 
For neutron-rich Ne isotopes $^{24-32}$Ne, 
the deviation of the closed circles from the stable-nucleus (dotted) line 
stems from density-broadening effects due to weak binding, although 
for $^{20-23}$Ne the deviation is originated 
in density-broadening effects due to deformation. 
The deviation of the open circles from the stable-nucleus line represents 
the contribution of $U_0$ to the density-broadening effects, and 
the deviation of the closed circles from the open ones does 
the contribution of $U_1$ to the density-broadening effects. 
The contribution of $U_1$ increases linearly as $N-Z$ becomes large, 
as expected. Both the contributions are comparable for 
neutron-rich Ne isotopes $^{24-32}$Ne.

Finally, we test the reliability of the isovector component $U_1$ of 
the present folding model by analyzing the charge-exchange $(p,n)$ reaction 
to isobaric analog states (IAS). 
The proton- and neutron-channel potentials and the coupling potentials 
are calculated from the Lane potential Eq. \eqref{Eq:Lane-potential}; 
see Ref. \cite{Khoa:2007ia} for the details of formulation.

\begin{figure}[htbp]
\begin{center}
 \includegraphics[width=0.4\textwidth,clip]{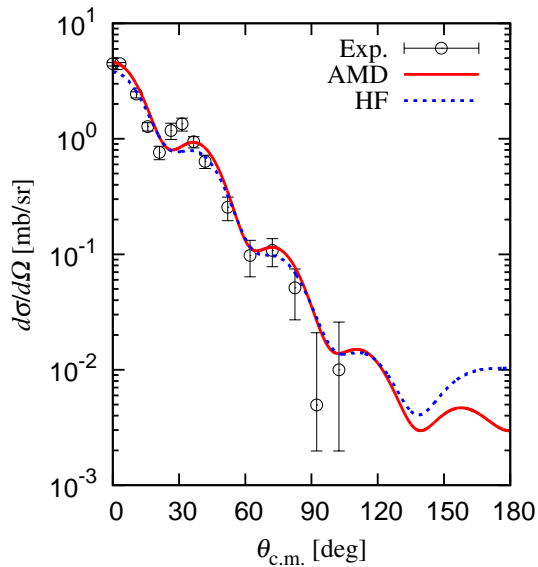}
 \caption{(Color online) 
Angular distribution of the differential cross section for $^{22}{\rm Ne}(p,n)^{22}{\rm F}_{\rm IAS}$ reaction at 35 MeV. The solid (dotted) line denotes the result of the folding model with the AMD (HF) density. 
The experimental data are taken from Ref. \cite{p-22Ne-IAS}.
  }
 \label{dcs-pNe22ias}
\end{center}
\end{figure}

Figure \ref{dcs-pNe22ias} shows the differential cross section 
of $^{22}{\rm Ne}(p,n)^{22}{\rm F}_{\rm IAS}$ reaction 
at 35 MeV. The solid and dotted lines represent the results 
of the folding model calculated with the AMD and HF densities, respectively. 
The two lines are close to each other at forward and intermediate 
angles $\theta_{\rm c.m.} \la 110^{\circ}$ 
where the experimental data \cite{p-22Ne-IAS} are available. The two lines well reproduce the data with no adjustable parameter, 
indicating that the present model is reliable also for the isovector component. Further analyses will be made in a forthcoming paper.

\section{Summary}
\label{Summary}

We have derived local microscopic optical potentials 
systematically for polarized proton scattering at 65~MeV, using 
the local-potential version of the Melbourne $g$-matrix folding model  
that was successful in reproducing nucleus-nucleus scattering~\cite{Min11,Min12,Sum12}. 
The local microscopic optical potentials for proton scattering 
well reproduce the experimental data on stable target nuclei systematically 
and have geometries similar to the phenomenological optical potentials such 
the Koning-Delaroche \cite{Koning-Delaroche} and the Dirac phenomenology \cite{Dirac1,Dirac2} one. 
Also for $p$+$^{6}$He scattering at 71 MeV, the $g$-matrix folding model 
yields better agreement than the $t$-matrix folding model. 
Therefore we can say that 
the local-potential version of the Melbourne $g$-matrix folding model is 
a reliable model to describe both nucleon-nucleus and nucleus-nucleus 
scattering.

The local microscopic optical potentials $U$ are also quite useful 
in many applications. 
Furthermore, the local nature makes it easy to clarify global properties 
of $U$ over both stable and unstable target nuclei. 
As target nuclei, in this paper we considered 
$^{6}$He and neutron-rich Ne isotopes 
in addition to stable nuclei of mass number $A=4$--$208$. 
In particular, it is interesting to compare properties of $U$ 
in $^{6}$He and $^{31}$Ne with general properties of $U$ 
in stable nuclei, since $^{6}$He is a typical two-neutron halo nucleus 
and $^{31}$Ne is a one-neutron halo nucleus with large deformation. 
Throughout the global analyses, we have clarified the $A$ and 
$N$ dependence of $U$. 
The target density is broadened in $^{6}$He by the weak-binding nature 
and in neutron-rich Ne isotopes by both the weak-binding nature 
and the nuclear deformation. 
The density broadening enlarges both the strength and the root-mean-square 
radius of the central part of $U$, but for the real spin-orbit part 
it enlarges the root-mean-square radius but weakens the strength. 
Density broadening effects due to weak binding and nuclear deformation are 
conspicuous particularly for halo nuclei such as $^{6}$He and $^{31}$Ne. 
The statement mentioned above is true also for $U$ 
at intermediate energies such as 200~MeV.

In proton scattering, target deformation not only broadens the spherical part 
of the folding potential but also enhances target excitations 
and back reactions to the elastic channel. 
For neutron-rich unstable target nuclei at 65 MeV, 
both the effects are appreciable, but density-broadening effects 
due to weak binding are much more important than the effects. 
For odd target nuclei with finite total spin, a reorientation 
of the target spin affects 
elastic scattering only at middle and backward angles, 
so the reorientation effects are small on reaction cross sections.

As mentioned above, density-broadening effects are quite important 
for neutron-rich nuclei. The effects are separated into 
the isoscalar and isovector components. Both the components yield 
comparable contributions to the density-broadening effects 
for neutron-rich nuclei. 
Finally, we tested the reliability of the isovector component 
of the present folding model by analyzing 
the charge-exchange $(p,n)$ reaction to isobaric analog states (IAS). 
The model well reproduces the experimental 
data on $^{22}{\rm Ne}(p,n)^{22}{\rm F}_{\rm IAS}$ 
reaction at 35 MeV. The present folding model is thus reliable also 
for the isovector component. 
Further analyses will be made for charge-exchange $(p,n)$ reactions 
in a forthcoming paper.

\vspace*{5mm}

\section*{Acknowledgments}

The authors thank T. Furumoto, T. Matsumoto, M.~Kohno, K. Ogata, 
T. Uesaka, and S. Sakaguchi for useful discussions.
This work is supported in part by Grant-in-Aid for Scientific Research
(No. 244137) 
from the Japan Society for the Promotion of Science (JSPS).
The numerical calculations of this work were performed
on the computing system in Research Institute
for Information Technology of Kyushu University.

\vspace*{5mm}

\appendix
\section{Definitions of quantities}
\label{Definitions}
The explicit forms of $\delta_0^{(\alpha)}(R,s)$ 
and $\rho_1^{(\alpha)}(R,s)$ are 
\bea
&& \delta_0^{(\alpha)}(R,s)\nonumber \\
&& ~~=\frac{1}{2}\int_{-1}^{+1} d\omega
  \frac{g^{\rm EX}_{{\rm LS},p\alpha}( s;\rho_{\alpha})}{x}
  \nonumber \\
&& ~~~~~\times \left\{\left. \frac{3}{k^{(\alpha)}_F(x) s}
		     j_1(k^{(\alpha)}_F(x) s) \frac{d}{dx}
		     \rho_{\alpha}(x) \right|_{x=\sqrt{R^2+s^2/4+R s
\omega }} \right. \nonumber \\
&& ~~~~~\left. +s\rho_{\alpha}(x) \frac{d}{dx} k^{(\alpha)}_F(x) \right|_{x=\sqrt{R^2+s^2/4+R s
\omega }} \nonumber \\
&& ~~~~~\left.\left. \times \frac{d}{dy}\left[\frac{3}{y}j_1(y)
					    \right] \right|_{y=k^{(\alpha)}_F(x) s} \right\}, 
\eea
and 
\bea
&& \rho_1^{(\alpha)}(R,s) \nonumber \\
&& ~~ =\frac{1}{2}\int_{-1}^{+1}d\omega ~\omega g^{\rm EX}_{{\rm
 LS},p\alpha}(s;\rho_{\alpha})\frac{3}{k^{(\alpha)}_F(x) s} \nonumber \\
&& ~~~~~ \left. \times j_1(k^{(\alpha)}_F(x) s)\rho_{\alpha}(y) \right|_{y=\sqrt{R^2+s^2/4+R s \omega}}, 
\eea
where 
\bea
&& k^{(\alpha)}_F = (3 \pi^2 \rho_{\alpha})^{1/3}, \\
&& g^{\rm DR,EX}_{{\rm LS},pp} (s;\rho_p)= g^{11}_{\rm LS},\\
&& g^{\rm DR,EX}_{{\rm LS},pn} (s;\rho_n)= \frac{1}{2}\left( \pm g^{10}_{\rm LS} + g^{11}_{\rm LS}  \right) .
\eea


\end{document}